\documentclass[tightenlines,
twocolumn,prl,notitlepage,nofootinbib,preprintnumbers,longbibliography]{revtex4-1}
\usepackage{graphicx}
\usepackage{amsmath}
\usepackage{amsfonts}
\usepackage{amssymb}
\usepackage{float}
\usepackage[colorlinks]{hyperref}
\usepackage{multirow}
\usepackage{bbold}
\usepackage{soul}
\usepackage{makecell}
\usepackage{minibox}
\usepackage{soul}
\usepackage{fancyhdr}
\usepackage{mathtools}
\usepackage{amsthm}
\usepackage{tikz}
\usepackage{placeins}
\usepackage[percent]{overpic}
\usepackage{siunitx} % to use floating-point alignment in tables (option "S")
\usepackage{subfigure}
\usepackage{tabularx}

\usepackage{pgfplots}
\pgfplotsset{compat=1.18}
\usepackage{enumitem,amssymb}
\newlist{todolist}{itemize}{2}
\setlist[todolist]{label=$\square$}
\usepackage{pifont}
\begin{document}
\title{Operational characterization of LAPPD Generation 2: charge sharing, delayed pulses, and dark-count behavior}
%Characterization of Charge-Sharing and Cross-Talk in LAPPD Generation 2}
\author{S.-W. Stradleigh}
\email{sstradle@ur.rochester.edu}
\affiliation{University of Rochester}
\author{J.A. Foot} \affiliation{University of California, Merced}
\author{R. Zhang} \affiliation{University of California, Merced}

\author{V.A. Li}%
\affiliation{Lawrence Livermore National Laboratory}
\date{\today}

\begin{abstract}
We present a study of charge sharing and electronic cross-talk in second-generation Large-Area Picosecond Photodetectors (LAPPD Gen 2). The LAPPD is a vacuum-based device consisting of a photocathode, two microchannel plates, and a resistive anode that capacitively couples to an 8 $\times$ 8 pixelated readout board (25.4 mm $\times$ 25.4 mm pixel area). Using a picosecond pulsed laser, we measure signal distributions across the resistive anode and quantify coupling between target and neighboring pixels. We further examine the relationship between dark-count rate and LAPPD voltage settings, identifying decay behavior characterized by fast, intermediate, and slow relaxation timescales. We additionally observe the LAPPD behaving as a resonant cavity by injecting electrical pulses into the readout board. To further interpret observed signals, we develop a pulse-classification method and identify additional features at approximately 60 ns and 110 ns. Finally, we implement a first-principles Monte Carlo simulation to model the radial and temporal distributions of observed signals, including contributions from electron backscatter and potential ion afterpulsing. The simulation shows reasonable agreement with the experimentally derived pulse classifications.

\end{abstract}
\maketitle

\section{Introduction}

\begin{figure} [hbt!]
    \centering
    \begin{overpic}[width=1\linewidth]{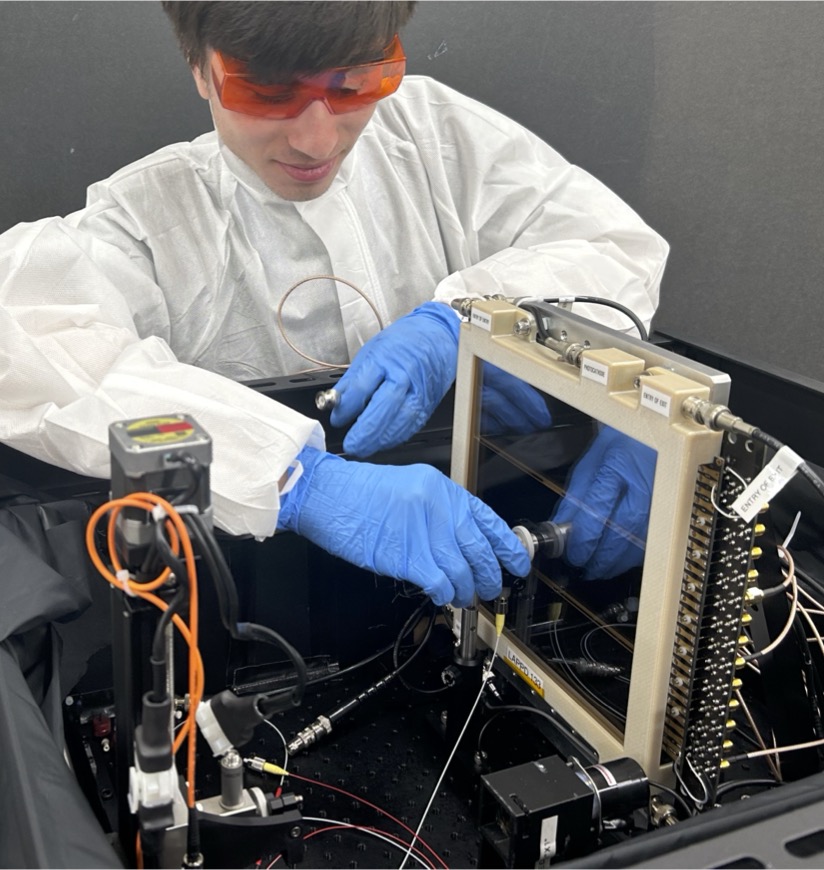}
        \put(56,50){\fcolorbox{red}{white!30}{\parbox{40pt}{\centering \textcolor{black}LAPPD}}}

        \put(62,37){\color{red}\vector(1,0){10}}
        \put(72,35){\fcolorbox{red}{white!30}{\parbox{45pt}{\centering \textcolor{black}Fixed collimator}}}

        \put(58,10){\fcolorbox{red}{white!30}{\parbox{20pt}{\centering 
        \textcolor{black}PMT}}}

        \put(25,8){\color{red}\vector(0,1){48}}
        \put(15,60){\fcolorbox{red}{white!30}{\parbox{50pt}{\centering \textcolor{black}Collimator on $xy$ stage}}}
    \end{overpic}
    \caption{The setup used in this study. The default LAPPD readout board consists of an 8 $\times$ 8 pixel pad array to act as the pickup electrode, which feeds signal to 64 SMA outputs, shown in this photograph. The pixels are 25.4~mm $\times$ 25.4~mm in size. The laser beam passing through a fiber and a collimator is being aligned with the center of pad E5 in this case. The $XY$ alignment stage with a second collimator is visible on the left.} 

    \label{fig:ShangWen_LAPPD_setup}
\end{figure}
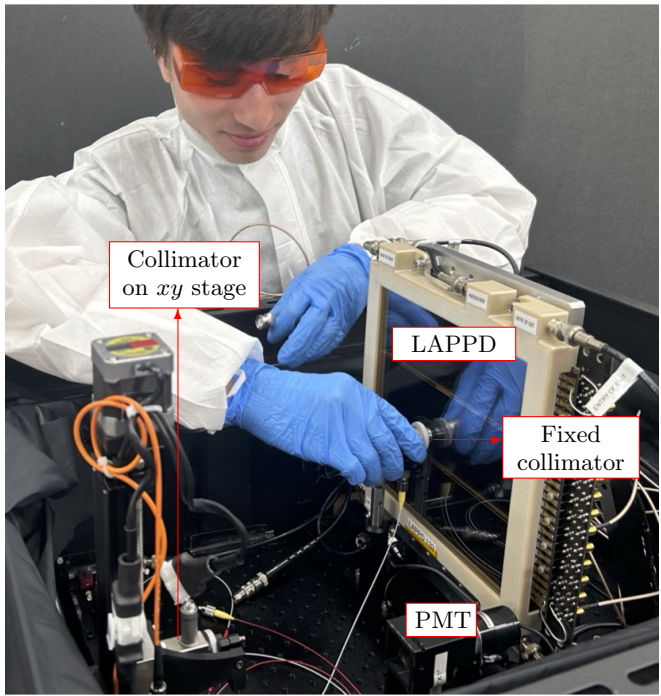

In this study, we aim to understand how signals spread between pixels in a microchannel plate LAPPD with a pixelated readout (generation 2), shown in Fig.~\ref{fig:ShangWen_LAPPD_setup}. To study this, we use a laser to scan across individual pixels and record the signal responses from the main pixel and its neighboring pixels. By analyzing how the signal distributes, we hope to better understand the cross talk behavior, which is important for improving the accuracy of readout systems, especially in detectors that utilize fast timing. Earlier studies were examined to understand microchannel plate physics, with Wiley and Goodrich from 1962 providing one of the earliest descriptions of electron multiplication within continuous-dynode electron multipliers modeled in tubular form \cite{Goodrich1962}. Back then, the models used to study MCPs were much simpler than today. For example, they assumed that the electric field inside the channels was uniform. The main parameter that was focused upon was the gain, which tells you how many electrons come out for each one that enters. They used simple formulas such as G = $\delta^n$, where $\delta$ is the secondary electron emission coefficient and $n$ is the number of collisions inside the channel. This $n$ can be estimated using $n=L / (d \tan\theta)$ where $L$ is the channel length, $d$ is the diameter, and $\theta$ is the tilt angle of the channel. These basic equations helped guide MCP design in the early days, but they did not consider things like signal spreading between pixels in modern systems. Our work aims to fill that gap and support the development of future fast and precise MCP based detectors.

The advent of experimental neutrino physics has resulted in developments of materials and photonics devices with the aim of increasing event detection probability and the signal-to-noise ratio. One such development is the water-based liquid scintillator (WbLS) with the idea of optimizing the ratio between scintillator and Cherenkov photon emissions to maximize the available information from a neutrino event~\cite{Yeh:2011zz}. Separating and characterizing scintillation from Cherenkov photons requires next-generation photodetectors with high temporal resolutions, in addition to a large surface detection area to maximize the probability of recording such an event~\cite{Aberle_2014}. For these reasons, the Gen-2 LAPPD was developed and is the focus of this study to discuss and understand its implementation within next-generation neutrino observatories from experiments like CHESS and the ANNIE group, with proposals for its use in DUNE under the THEIA detector~\cite{anniecollab2024,Caravaca_2020,Kaptanoglu_2022,Askins}.

\section{LAPPD Gen 2 and setup}

\begin{figure} [hbt!]
    \hspace*{1.2cm}
    \begin{overpic}[width=0.7\linewidth]%,grid,tics=5]
         {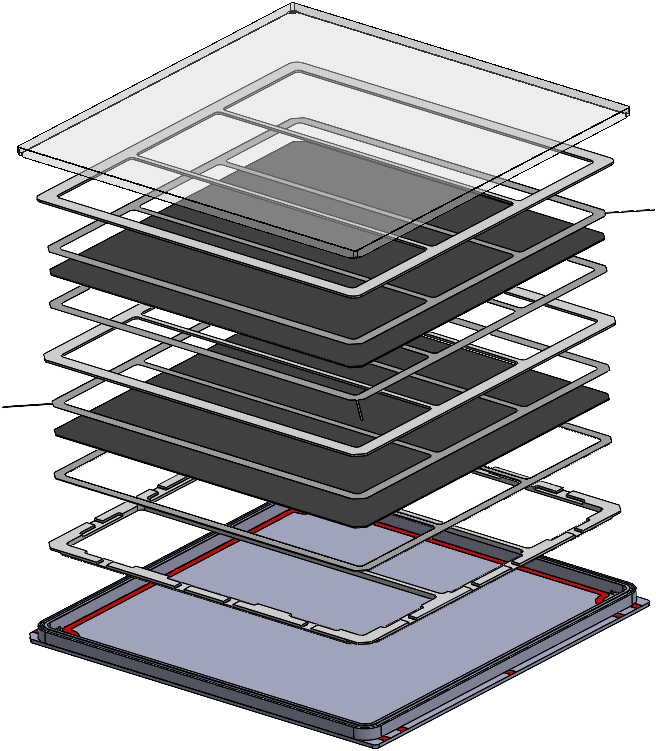}
        \put(0,80){\color{black}\vector(1,0){5}}
        \put(-28,78){Photocathode}
        \put(3, 64){\color{black}\vector(1,0){5}}
        \put(-18,62){Top MCP}
        \put(3, 42){\color{black}\vector(1,0){5}}
        \put(-25,40){Bottom MCP}
        \put(0,15){\color{black}\vector(1,0){5}}
        \put(-15, 14){Anode}
    \end{overpic}
    \caption{CAD rendering: exploded view of the LAPPD generation 2. Figure is from Incom used with permission.}
    \label{fig:LAPPD_CAD}
\end{figure}
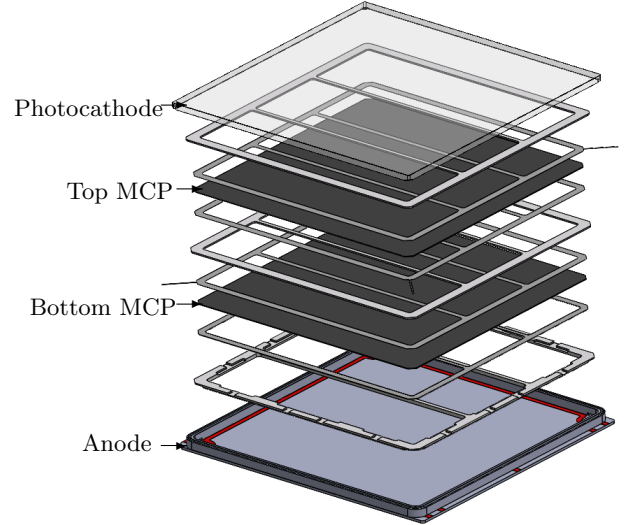

The second generation LAPPD (seen in Fig.~\ref{fig:ShangWen_LAPPD_setup}) is a multi-anode device with two micro-channel plates in a glass enclosure under vacuum, enabling its function as a photomultiplier tube (PMT). It is constructed with a 5.0-mm thick fused-silica glass window for light transmission with a measured active area of 195~mm $\times$ 195~mm for a minimum total effective area of 373~cm$^2$. The design additionally includes supporting edge frame rib-spacers to minimize dead space, keeping an active fraction of 97\% to enhance light collection efficiency.
This leads to a bialkali (Na$_2$KSb) photocathode material as seen in Fig.~\ref{fig:LAPPD_CAD}, which has an effective spectral response range of 160 to 650~nm photons, converting such photons to photoelectrons. It is maximally efficient for photons at a wavelength of 365~nm or less, with reasonable efficiency achieved at 405~nm, which is of interest for detecting Cherenkov emissions. 
There are two 203~mm $\times$ 203~mm MCPs with a pore size of 20~$\mu$m to generate secondary emissions for electron amplification after the photocathode. 
High voltage cables are connected to the photocathode, the entry of top MCP, exit of top MCP, entry of bottom MCP, exit of bottom MCP, and to the anode which is grounded. This generates an applied electric field that pulls photoelectrons from the photocathode towards MCPs and the resulting electron cloud to the resistive anode. The deposited charge is recorded by the resistive anode, which then induces a current that is capacitively coupled to the pickup electrodes, which is an array of conductors and a signal ground to create an external readout board. The default LAPPD readout board consists of an 8 $\times$ 8 pixel pad array to act as the pickup electrode, which feeds signal to 64 SMA outputs. Detailed component measurements are mentioned in Table~\ref{table:LAPPD_Dimensions}.

\begin{table}[h!]
\centering
\begin{tabular}{|l|S|}
\hline
\textbf{Component} & \textbf{Thickness (mm)} \\
\hline\hline
Window (fused silica) & 5.0 \\\hline
Top Spacer & 2.59 \\
Entry Top Shim & 0.25 \\
Entry MCP & 1.2 \\
Entry Bottom Shim & 0.05 \\\hline 
Middle Spacer & 1.14 \\\hline
Exit Top Shim & 0.05 \\
Exit MCP & 1.2 \\
Exit Bottom Shim & 0.38 \\
Bottom Spacer & 6.35 \\ \hline

Anode (ceramic) & 2.0 \\ \hline

Stack Height & 13.22 \\
Sidewall Height &  13.18\\ \hline

Total Height & 20.22 \\
\hline
\end{tabular}
\caption{Component dimensions in millimeters}
\label{table:LAPPD_Dimensions}
\end{table}

The experimental setup consists of an enclosed laser dark-box mounted on a mobile cart, which contains the LAPPD, high-voltage cables connected to the LAPPD, a voltage divider, laser fibers fed from the laser head, a X-Y motion slides with motion controller to direct a laser beam at a specified pixel, collimators, and neutral density filters. SMA output cables are fed out of the dark-box to connect to a data acquisition system. The cart itself contains the laser controller, a 450~nm laser head that is directed into laser fibers, a delay generator, and a high-voltage power supply.

We have used Tektronix MSO64B oscilloscope (at 25 GSa/s sampling rate) as a data acquision system.
We have automated data collection and processing using Tektronix Programming Manual.
A Linux machine was used to operate the oscilloscope and stepper motor controller using BASH scripting.
The algorithm runs in a loop, for each position the laser beam aligns to a spot determined by some interval position separation, and then the set of commands is send to the oscilloscope (recording data from both the target and the neighbor/s pixels). This method is used to position the laser without requiring the need to access the darkbox, however we do not perform any specific raster-scan data acquisition in this study.

\section{Electronic cross-talk and charge-sharing}
Within electronics, signal transmission on one channel may produce signal on another channel through capacitive, inductive, conductive coupling, or from electromagnetic interference. The signal that is observable on the secondary channels due to this phenomenon is known as cross-talk and typically exhibits negative-polarity in pixelated photodetectors as the waveform starts. Alternatively, charge-sharing, as the name implies, can induce a signal on neighboring channels through the diffusion of electrical charge from the main signal in the primary channel.

\begin{figure} [hbt!]
    \centering
    \includegraphics[width=1\linewidth]{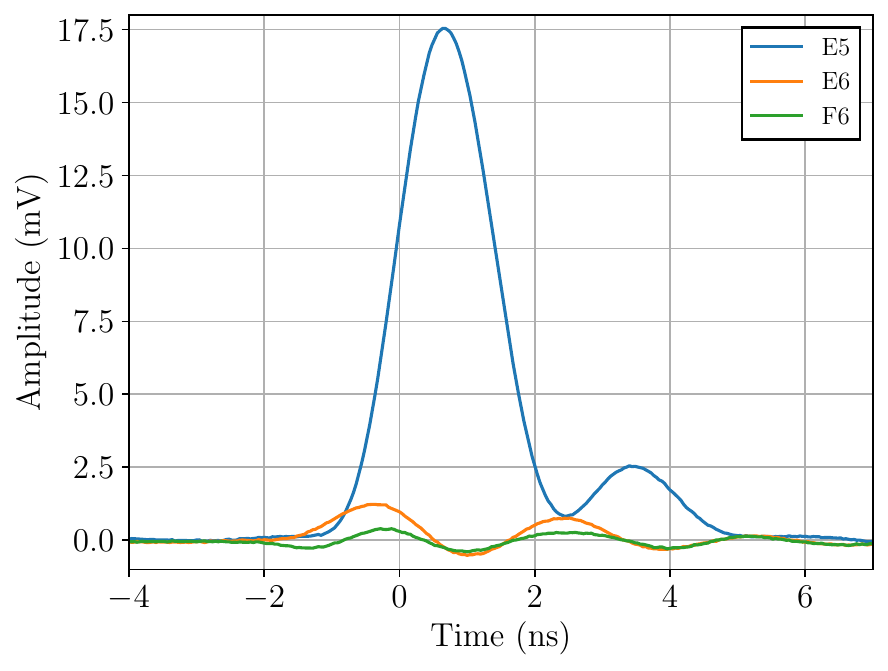}
    \caption{Example of cross-talk in neighboring pixels (E6/F6) from signal on the target pixel (E5). Note that the cross-talk signal in pixel F6 appears in the shape of an underdamped harmonic oscillator, similar to work found in Bhattacharya\cite{Bhattacharya:2023nmj}. } 
    \label{fig:E5_CrosstalkExample}
\end{figure}

Previous work by the authors and other studies involving LAPPD signal characterization  \cite{Slava2025_RSI,Bhattacharya:2023nmj} discovered observable signal spread between pixels understood as cross-talk and charge-sharing when reading multiple pixel channels simultaneously. Figure \ref{fig:E5_CrosstalkExample} shows a visual observation of electronic cross-talk waveforms between different pixel channels, with the signal on target E5 pixel creating a notable cross-talk on pixel F6 and charge-sharing signal to E6.

\begin{figure}
\setlength{\unitlength}{.2\linewidth}
    \centering
\begin{picture}(5,2)
\put(1.3,1.65){\color{blue}\vector(1,0){0.3}}
\put(2.32,1.65)
{\color{blue}\line(1,0){1.2}}
\put(2.32,0.5)
{\color{blue}\line(1,0){1.2}}
\put(3.515,0.5){\color{blue}\vector(0,1){0.4}}
\put(3.515,1.65){\color{blue}\vector(0,-1){0.4}}
\put(0.5,1.65){\fbox{LASER}}

\put(1.7,1.3){\rotatebox{90}{\fbox{\shortstack{Target\\Pixel}}}}

\put(1.7,0){\rotatebox{90}{\fbox{\shortstack{Neighbor\\Pixel}}}}
\put(3,1.03){{\fbox{Oscilloscope}}}
\put(1,1){Cross-talk}
\put(1.95,1.25){\color{black}\vector(0,-1){0.35}}
\end{picture}
    \caption{The setup to investigate cross-talk in the LAPPD, with cross-talk from the signal on Pixel E5 generating signal in neighboring pixels}
    \label{fig:Crosstalk_Setup}
\end{figure}
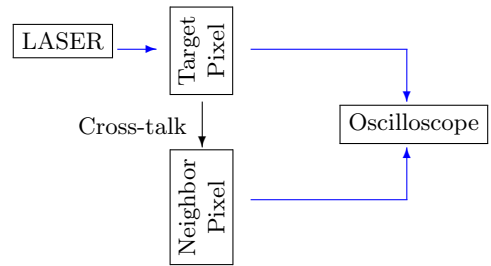

To characterize the behavior of cross-talk and the charge-sharing effects that occur on neighboring pixels, a laser source was aimed at a specific pixel, in our case pixel E5, on the readout board to measure a traditional photoelectron pulse. SMA output connectors to the target and neighboring pixels were then fed into the oscilloscope, where the cross-talk response can be measured relative to the signal output. This setup is depicted in Figure \ref{fig:Crosstalk_Setup}, and shows how the laser signal on the target pixel generates cross-talk on neighboring pixels that can then be seen on a readout system.
\begin{figure}
    \centering
    \begin{tikzpicture}
  \node[draw, shape=rectangle, minimum width=14pt, minimum height=14pt, fill=red!60] at (0,0.0) {};
  \node[draw, shape=rectangle, minimum width=14pt, minimum height=14pt, fill=yellow!60] at (0,0.6) {};
  \node[draw, shape=rectangle, minimum width=14pt, minimum height=14pt, fill=yellow!60] at (0.6,0.6) {};
  \node[draw, shape=rectangle, minimum width=14pt, minimum height=14pt, fill=yellow!60] at (0.6,-0.6) {};
  \node[draw, shape=rectangle, minimum width=14pt, minimum height=14pt, fill=yellow!60] at (-0.6,-0.6) {};
  \node[draw, shape=rectangle, minimum width=14pt, minimum height=14pt, fill=yellow!60] at (-0.6,0.6) {};
  \node[draw, shape=rectangle, minimum width=14pt, minimum height=14pt, fill=yellow!60] at (0,-0.6) {};
  \node[draw, shape=rectangle, minimum width=14pt, minimum height=14pt, fill=yellow!60] at (0.6,0) {};
  \node[draw, shape=rectangle, minimum width=14pt, minimum height=14pt, fill=yellow!60] at (-0.6,0) {};
  
  \node[node font=\scriptsize] at (0,0.6) {5\%};
  \node[node font=\scriptsize] at (0,0.01) {100};
  \node[node font=\scriptsize] at (0.01,-0.6) {8\%};
  \node[node font=\scriptsize] at (-0.6,-0.6) {2\%};
  \node[node font=\scriptsize] at (-0.6,0.01) {6\%};
  \node[draw, shape=rectangle, minimum width=14pt, minimum height=14pt, fill=yellow!60] at (-1.2,0) {};
  \node[node font=\scriptsize] at (-1.2,0.01) {.5\%};
  \node[draw, shape=rectangle, minimum width=14pt, minimum height=14pt, fill=yellow!60] at (-1.8,0) {};
  \node[node font=\scriptsize] at (-1.8,0.01) {.5\%};
  \node[draw, shape=rectangle, minimum width=14pt, minimum height=14pt, fill=yellow!60] at (-2.4,0) {};
  \node[node font=\scriptsize] at (-2.4,0.01) {.5\%};
  \node[node font=\scriptsize] at (-0.6,0.59) {3\%};
  \node at (-0.6,3) {4};
  \node at (-1.2,3) {3};
  \node at (-1.8,3) {2};
  \node at (-2.4,3) {1};
  \node at (-3,2.4) {A};
  \node at (-3,1.8) {B};
  \node at (-3,1.2) {C};
  \node at (-3,0.6) {D};
  \node at (-3,0) {E};
  \node at (-3,-0.6) {F};
  \node at (-3,-1.2) {G};
  \node at (-3,-1.8) {H};
  \node at (0,3) {5};
  \node at (0.6,3) {6};
  \node at (1.2,3) {7};
  \node at (1.8,3) {8};
  \node[node font=\scriptsize] at (0.6,0.6) {1\%};
  \node[node font=\scriptsize] at (0.6,0) {7\%};
  \node[node font=\scriptsize] at (0.6,-0.6) {2\%};

  \node[draw, shape=rectangle, minimum width=14pt, minimum height=14pt, fill=orange!60] at (-1.2,0.6) {};
  \node[draw, shape=rectangle, minimum width=14pt, minimum height=14pt, fill=orange!60] at (-1.8,0.6) {};
  \node[draw, shape=rectangle, minimum width=14pt, minimum height=14pt, fill=orange!60] at (1.8,0) {};
  \node[draw, shape=rectangle, minimum width=14pt, minimum height=14pt, fill=orange!60] at (1.2,0) {};
  \node[draw, shape=rectangle, minimum width=14pt, minimum height=14pt, fill=orange!60] at (1.8,-0.62) {};
  \node[draw, shape=rectangle, minimum width=14pt, minimum height=14pt, fill=orange!60] at (1.2,-0.62) {};
  \node[draw, shape=rectangle, minimum width=14pt, minimum height=14pt, fill=orange!60] at (1.8,-1.2) {};
  \node[draw, shape=rectangle, minimum width=14pt, minimum height=14pt, fill=orange!60] at (1.2,-1.2) {};
  \node[draw, shape=rectangle, minimum width=14pt, minimum height=14pt, fill=orange!60] at (0.6,-1.2) {};
  \node[draw, shape=rectangle, minimum width=14pt, minimum height=14pt, fill=orange!60] at (0,-1.2) {};
  \node[draw, shape=rectangle, minimum width=14pt, minimum height=14pt, fill=orange!60] at (-0.6,-1.2) {};
  \node[draw, shape=rectangle, minimum width=14pt, minimum height=14pt, fill=orange!60] at (-1.2,-1.2) {};
  \node[draw, shape=rectangle, minimum width=14pt, minimum height=14pt, fill=orange!60] at (-1.8,-1.2) {};
  \node[draw, shape=rectangle, minimum width=14pt, minimum height=14pt, fill=orange!60] at (-2.4,-1.2) {};
  \node[draw, shape=rectangle, minimum width=14pt, minimum height=14pt, fill=orange!60] at (1.8,-1.8) {};
  \node[draw, shape=rectangle, minimum width=14pt, minimum height=14pt, fill=orange!60] at (1.2,-1.8) {};
  \node[draw, shape=rectangle, minimum width=14pt, minimum height=14pt, fill=orange!60] at (0.6,-1.8) {};
  \node[draw, shape=rectangle, minimum width=14pt, minimum height=14pt, fill=orange!60] at (0,-1.8) {};
  \node[draw, shape=rectangle, minimum width=14pt, minimum height=14pt, fill=orange!60] at (-0.6,-1.8) {};
  \node[draw, shape=rectangle, minimum width=14pt, minimum height=14pt, fill=orange!60] at (-1.2,-1.8) {};
  \node[draw, shape=rectangle, minimum width=14pt, minimum height=14pt, fill=orange!60] at (-1.8,-1.8) {};
  \node[draw, shape=rectangle, minimum width=14pt, minimum height=14pt, fill=orange!60] at (-2.4,-1.8) {};
  \node[draw, shape=rectangle, minimum width=14pt, minimum height=14pt, fill=orange!60] at (1.8,1.2) {};
  \node[draw, shape=rectangle, minimum width=14pt, minimum height=14pt, fill=orange!60] at (1.2,1.2) {};
  \node[draw, shape=rectangle, minimum width=14pt, minimum height=14pt, fill=orange!60] at (0.6,1.2) {};
  \node[draw, shape=rectangle, minimum width=14pt, minimum height=14pt, fill=orange!60] at (0,1.2) {};
  \node[draw, shape=rectangle, minimum width=14pt, minimum height=14pt, fill=orange!60] at (-0.6,1.2) {};
  \node[draw, shape=rectangle, minimum width=14pt, minimum height=14pt, fill=orange!60] at (-1.2,1.2) {};
  \node[draw, shape=rectangle, minimum width=14pt, minimum height=14pt, fill=orange!60] at (-1.8,1.2) {};
  \node[draw, shape=rectangle, minimum width=14pt, minimum height=14pt, fill=orange!60] at (-2.4,1.2) {};
  \node[draw, shape=rectangle, minimum width=14pt, minimum height=14pt, fill=orange!60] at (1.8,1.8) {};
  \node[draw, shape=rectangle, minimum width=14pt, minimum height=14pt, fill=orange!60] at (1.2,1.8) {};
  \node[draw, shape=rectangle, minimum width=14pt, minimum height=14pt, fill=orange!60] at (0.6,1.8) {};
  \node[draw, shape=rectangle, minimum width=14pt, minimum height=14pt, fill=orange!60] at (0,1.8) {};
  \node[draw, shape=rectangle, minimum width=14pt, minimum height=14pt, fill=orange!60] at (-0.6,1.8) {};
  \node[draw, shape=rectangle, minimum width=14pt, minimum height=14pt, fill=orange!60] at (-1.2,1.8) {};
  \node[draw, shape=rectangle, minimum width=14pt, minimum height=14pt, fill=orange!60] at (-1.8,1.8) {};
  \node[draw, shape=rectangle, minimum width=14pt, minimum height=14pt, fill=orange!60] at (-2.4,1.8) {};
  \node[draw, shape=rectangle, minimum width=14pt, minimum height=14pt, fill=orange!60] at (1.8,2.4) {};
  \node[draw, shape=rectangle, minimum width=14pt, minimum height=14pt, fill=orange!60] at (1.2,2.4) {};
  \node[draw, shape=rectangle, minimum width=14pt, minimum height=14pt, fill=orange!60] at (0.6,2.4) {};
  \node[draw, shape=rectangle, minimum width=14pt, minimum height=14pt, fill=orange!60] at (0,2.4) {};
  \node[draw, shape=rectangle, minimum width=14pt, minimum height=14pt, fill=orange!60] at (-0.6,2.4) {};
  \node[draw, shape=rectangle, minimum width=14pt, minimum height=14pt, fill=orange!60] at (-1.2,2.4) {};
  \node[draw, shape=rectangle, minimum width=14pt, minimum height=14pt, fill=orange!60] at (-1.8,2.4) {};
  \node[draw, shape=rectangle, minimum width=14pt, minimum height=14pt, fill=orange!60] at (-2.4,2.4) {};
  \node[draw, shape=rectangle, minimum width=14pt, minimum height=14pt, fill=orange!60] at (1.8,0.6) {};
  \node[draw, shape=rectangle, minimum width=14pt, minimum height=14pt, fill=orange!60] at (1.2,0.6) {};
  \node[draw, shape=rectangle, minimum width=14pt, minimum height=14pt, fill=orange!60] at (-2.4,0.6) {};
  \node[draw, shape=rectangle, minimum width=14pt, minimum height=14pt, fill=orange!60] at (-1.2,-0.6) {};
  \node[draw, shape=rectangle, minimum width=14pt, minimum height=14pt, fill=orange!60] at (-1.8,-0.6) {};
  \node[draw, shape=rectangle, minimum width=14pt, minimum height=14pt, fill=orange!60] at (-2.4,-0.6) {};
\end{tikzpicture}
    \caption{A diagram of the pixelated readout board displaying the signal charge on each pixel relative to the signal waveform. The laser targets pixel E5 (colored red), while cross-talk signal is read out by the neighboring pixels (colored yellow).}
    \label{fig:LAPPDCrossTalk}
\end{figure}
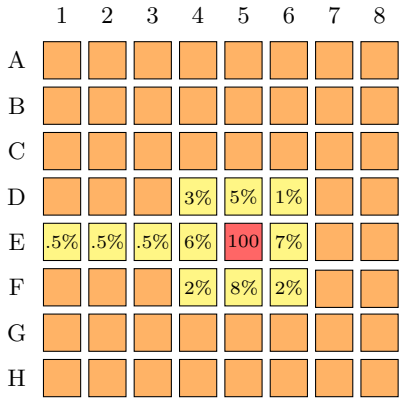

To analyze the cross-talk, we integrated the area under the photoelectron waveform signal determine its total charge. We compared the integrated charge of the cross-talk on neighboring pixels to the source signal as a reference to figure out the amount of charge-sharing that was occurring. Figure \ref{fig:LAPPDCrossTalk} shows the measured cross-talk charge relative to the source waveform. Most notably, we find that charge-sharing is stronger on the perpendicular direct-neighboring pixels compared to the diagonal/off-diagonal pixels, and that cross-talk falls below 1\% beyond 1-pixel neighbor. 

We note that this cross-talk is intrinsic to the LAPPD by design due to the capacitive coupling of the anode onto the pixelated readout grid. While cross-talk is traditionally viewed as an undesired effect in electronics, it in fact can be advantageous in the LAPPD for event reconstruction. Depending on the precise location of a photon that enters the photocathode to strike a targeted pixel, the cross-talk response may change even if the primary signal in the target pixel remains the same. This response would enable a user to perform an event reconstruction of the photon from its source.

\section{LAPPD As a Resonant Cavity}

\begin{figure}[ht]
\setlength{\unitlength}{.2\linewidth}
%\fbox{
\begin{picture}(4,1)

\put(0.6,0.05){\color{blue}\vector(1,0){.4}}
\put(2.4,0.05){\color{blue}\vector(1,0){.7}}
\put(3.6,0.05){\color{blue}\vector(1,0){.4}}

\put(0,0){\fbox{AFG}}
\put(1,-0.1){\fbox{\shortstack{Target Pixel\\(SMA Output)}}}
\put(3.1,-0.4){\rotatebox{90}{\fbox{\shortstack{Neighbor\\Pixel}}}}
\put(4,-0.5){\rotatebox{90}{\fbox{Oscilloscope}}}
\end{picture}
%} % end for \fbox command to outline boundaries
\vspace{1cm}
\caption{The schematic of the setup to study cross-talk resonance. The AFG plugs into the SMA output of the LAPPD, feeding signal into the target pixel that then drives a measurable cross-talk on the neighboring pixels.} 
\label{fig_splitter_dia}
\end{figure}
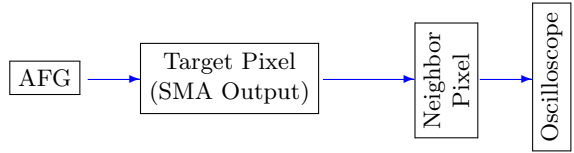

To study effect of cross talk and resonance in the LAPPD, we inject electrical sine-wave pulses into the target pixel pad using a signal generator and read out the neighboring pixels with an oscilloscope to measure cross talk characteristics. This setup is illustrated in Fig.~\ref{fig_splitter_dia}.

\begin{figure}
    \centering
    \includegraphics[width=1
    \linewidth]{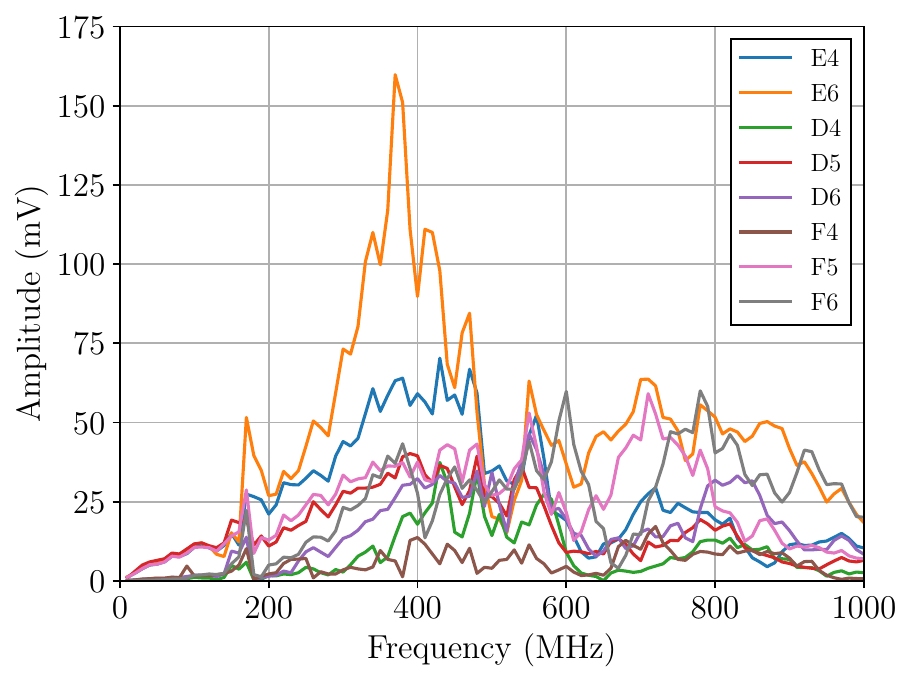}
    \caption{Cross-talk amplitude response as a function of sinusoidal RF input to the E5 target pixel. We note that there are resonant peaks for all pixels at 180 MHz and 550 MHz, while 350-450 MHz and 750-800 MHz have pixel offsets.}
    \label{fig:Resonance}
\end{figure}

We find that the cross talk signal amplitude has a dependence on the frequency of the injected sine-wave signal as seen in Fig. \ref{fig:Resonance}. While we do not fully understand its correlation with the LAPPD's electrical cross-talk, we note from these results that the LAPPD may be susceptible to receiving radio-frequency (RF) signals. If RF signals do interfere with LAPPD readout, then the exact positioning of the LAPPD in the setup may be of concern in experimental setups to avoid this interference.

\section{LAPPD Muon Detection}

\begin{figure}
    \centering
    \includegraphics[width=1\linewidth]{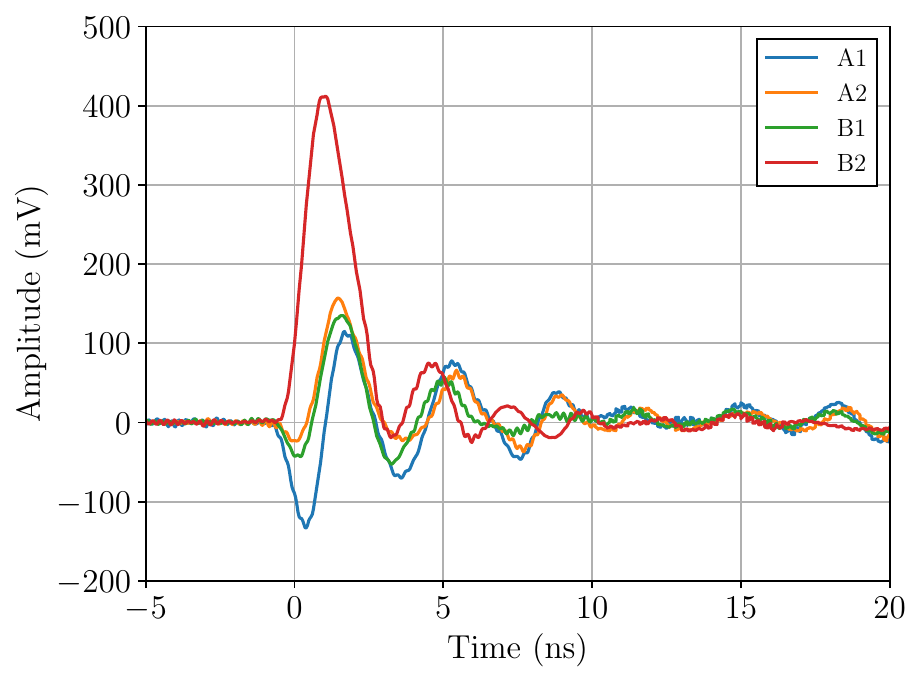}
    \caption{Oscilloscope trace recorded on neighboring pads A1, A2, B1, and B2 for a representative muon event. The negative-going onset observed in three channels indicates significant cross-talk (trigger is set on B2 signal at 100 mV).}
    \label{fig:muonsignal}
\end{figure}

During operation of the LAPPD, we observed an unusually large signal that would be regularly read by the oscilloscopes on all pixels. On nearly all pixels, it would display cross-talk like behavior with a large "ringing" of the waveform that dampened over a ~100~ns time threshold. 
We observed large, spatially distributed events at a rate of a few hertz. Based on their amplitude and occurrence rate, these events are consistent with candidate cosmic-ray interactions, although no external coincidence measurement was performed in the present study.
One such example of a large signal event is seen in Figure~\ref{fig:muonsignal}.

Investigation of the cross-talk signal revealed that this signal occurs on all pixel pads simultaneously, indicating that the source must be a highly charged event striking the LAPPD at some pixel. We were able to isolate and observe the source signal on a given target pixel in junction with the observed cross-talk on other neighboring pixels. The source signal has the characteristic photoelectron waveform with a maximum peak of ~700-800 mV, along with other peaks delayed in time. We observed these events occurring at a rate of $\sim$1-2 Hz.

We note that in theory, the rule of thumb for the flux of a $\sim$9 GeV muon at sea-level is 1 muon per cm$^2$ per minute with a sky-facing orientation (90-degree angle, zenith from the horizon), assuming flat terrain symmetry and clear weather. If we extend this rule of thumb to the LAPPD, each pixel pad has an area of $\sim$1 cm$^2$. Doing the straightforward multiplication with 64 pads in our LAPPD gives us an expected rate of $\sim$1 muon per second, which aligns with our observations. While there are computational methods that can more precisely predict the expected muon flux at our setup location, such work is outside the scope of this paper.

The muon, due to its highly charged nature and decay near the surface, would explain why the observed waveform has a large peak along with the delayed peaks. The delayed peaks are explained by the charge being deposited within the MCPs and their geometry, where backscattering and kinematics causes a low but significant number of electrons to have a delayed travel time onto the anode. We note that operation of the LAPPD for general experiments would require calibration in the data acquisition methodology to filter out the muon signal as potentially unwanted background and cross-talk noise.

\section{Study on high dark count rate}
We noticed that when we experimented with different high-voltage settings for the LAPPD, the behavior of the dark count rate would increase when a few distinct settings were applied. To investigate the systematic cause of this behavior, we looked at changing the voltage setting between each high-voltage channel. This would allow us to determine the relationship between each channels and their specific voltage setting with respect to changes in the dark-count rate.
To accommodate the current measurement during the high voltage stability test, the four wires connected to the LAPPD were temporarily unsolder, and one of them was routed through a digital ammeter. This lets us monitor for current instabilities associated with the voltage setting, even if we are operating in safe voltage parameters according to the manufacturer Incom.

\begin{figure}[H]
    \centering
    \includegraphics[width=1\linewidth]{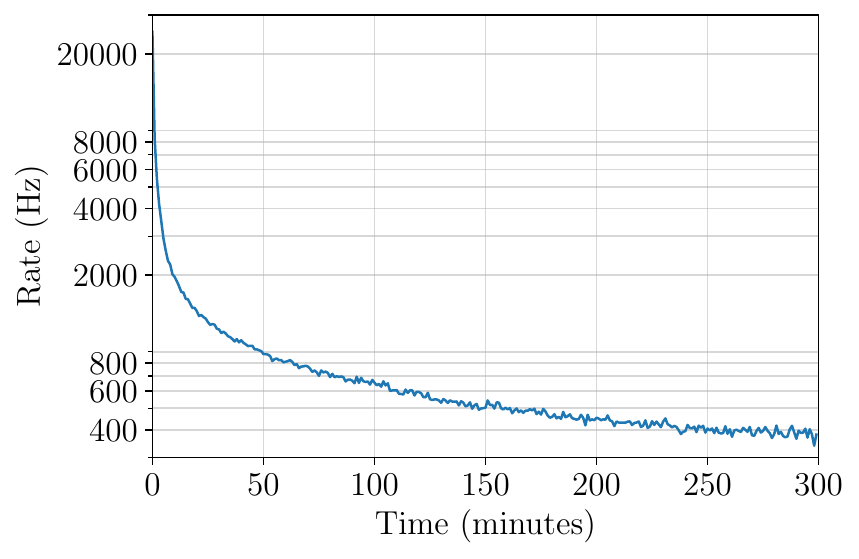}
    \caption{Dark count rate as a function of time after increasing the top MCP voltages. The decay is modeled with a tri-exponential function, revealing fast, intermediate, and slow relaxation components.}
    \label{fig:decay}
\end{figure}

The voltage settings were intentionally adjusted from (2200–2000–1200–1075–200) to (2200–2150–1275–1075–200) to test the stability of the dark count rate under the maximum operating conditions recommended by the LAPPD manufacturer. In this adjustment, the entry and exit voltages of the top MCP were increased and the new voltage configuration was kept constant during subsequent data acquisition. During measurement, the lights were turned off to avoid exposing the LAPPD to ambient light. On average, 31 measurements of the dark count rate were recorded every minute (within a 15 second window inside each minute) for a total duration of 300 minutes. For each minute, the mean of the 31 measurements was calculated and used as a single data point to produce the plot in Figure \ref{fig:decay}.

\begin{figure*} [hbt!]
    \centering
    \includegraphics[width=1.\linewidth]{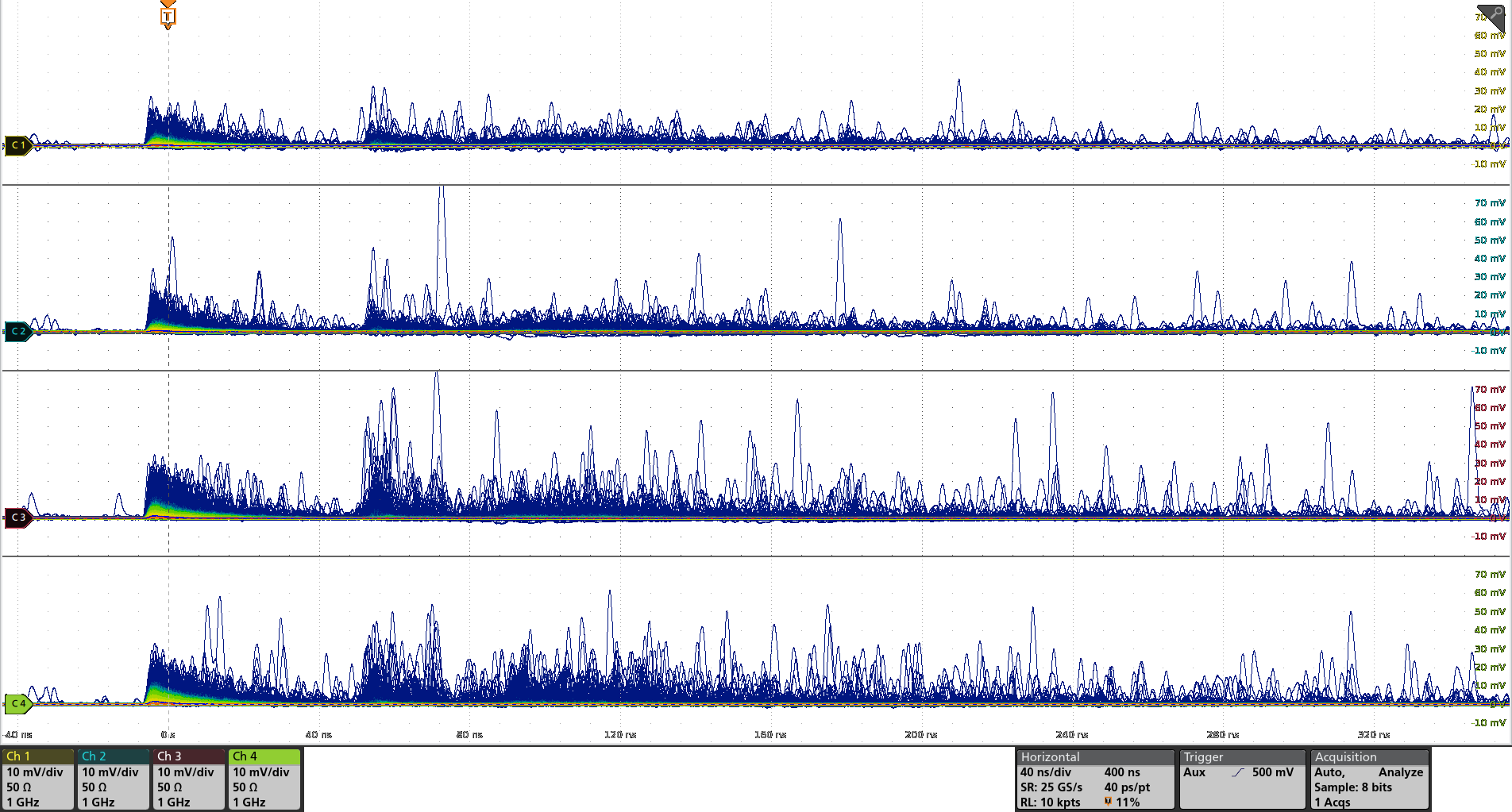}
    \caption{Oscilloscope Traces 10,000 waveforms per channel overplayed. Trigger on auxiliary input of the oscilloscope (coming from the delay generator which also triggers the laser controller). Noticeable cluster can be observed at around 60 ns and above due to ion feedback. Channels on this oscilloscope capture correspond to the following LAPPD pads: Ch1 --- A1; Ch2 --- A2; Ch3 --- B1; Ch4 --- B2.} %%%timing could differ from cable length possibly? -sw
    \label{fig:FastFrame}
\end{figure*}

\begin{equation}
\begin{aligned}
R(t) =\;& 361.5 
+ 1.92 \times 10^{4} e^{-t/0.80} \\
&+ 2.90 \times 10^{3} e^{-t/7.79} \\
&+ 9.92 \times 10^{2} e^{-t/78.7}.
\end{aligned}
\label{eq:triexp}
\end{equation}

Following the voltage change, an instantaneous increase in the dark count rate was observed, rising from approximately 200 Hz to 28 kHz. Notably, no such dark count spike was observed when the entry and exit voltages of MCP2 were increased to their maximum allowable values, indicating that the observed instability is specifically associated with the voltage configuration of the top MCP. This increase was accompanied by a spike in the measured current. The behavior is reproducible, with similar trends and values observed each time the setup is repeated. Over time, the LAPPD gradually returns to its normal dark count rate ( about 300 Hz) within about half a day for this particular test. The long recovery time suggests that the observed behavior is not dominated by a purely electronic transient but rather by an internal recovery process within the detector.\\
\indent To analyze the decay behavior following the dark count spike, a tri-exponential model \eqref{eq:triexp} was employed. This model introduces three characteristic time scales corresponding to a fast initial decay, an intermediate relaxation process, and a slow long term decay. The presence of multiple decay components indicates that the observed relaxation is governed by several underlying physical processes acting on different timescales, rather than a single dominant mechanism.\\ %Can we make a statement on what these mechanisms may be?
%The physics behind these events are not entirely clear (to the authors)
The decay behavior suggests that multiple mechanisms may influence the dark count rate in the LAPPD. We hypothesize that one such potential mechanism may be from temperature heating in the LAPPD, causing thermal emissions from the photocathode, microchannel plates, and ions. The change in the voltage setting could also have caused field emission, where electrons are emitted from the MCPs by the electric field. Similar behavior is also observed and well documented in photomultiplier tubes, where thermionic electrons are spontaneously emitted by the photocathode (even at room temperature) and field emissions occur from the dynodes at excessively high voltages, which corresponds with an abrupt increase in the dark current \cite{Bell1973,Hamamatsu2017}. We have no definitive knowledge of the physical mechanisms that cause all three decay characteristics and hope that future studies may be able to clarify our observations. 

To further investigate signal and dark count readout on the LAPPD, we directed a laser to measure signal out of the pixels A1, A2, B1, B2, and used the FastFrame acquisition readout mode capability of the Tektronix oscilloscope. This mode changes the trigger setting to a burst rate of $\geq5\times10^6$ frames (acquisitions) per second and overlays the waveforms together on a digital display, allowing for the collection of all relevant signals that can be detected for readout within seconds. This has the immediate advantage of revealing rare trigger events that may be hard to identify individually relative to the main trigger event, such as electronic back scatter or ion-feedback. One such raw capture is seen in Figure~\ref{fig:FastFrame}. One can observe that this FastFrame capture bins the traditional photoelectron waveforms around 0 ns, while an immediate cluster of additional waveforms are binned at 60 ns and above in each channel. This indicates that significant events occur after the trigger registers the initial photoelectron waveform. These afterpulses, being satellite pulses that follow a true signal pulse after some time delay, have been observed as a characteristic in PMTs \cite{Morton1967}. Within PMTs, short-delay (20-50 ns) pulses result from electrons elastically scattering on the dynode, while those with longer delays (100 ns-1 $\mu$s) come from ionized residual gas such as Helium due to the imperfect vacuum \cite{STAUBERT1970,HALL1973,AKCHURIN2007}. We hypothesize that our observations must be ion-feedback pulses and/or electron backscattering occurring in the LAPPD (which can be described as a multi-anode MCP-PMT, so the similarity to PMT afterpulsing is no coincidence) and discuss further investigations in our discussion of our 2D LAPPD simulation model.

\section*{Pulse Event Classification}

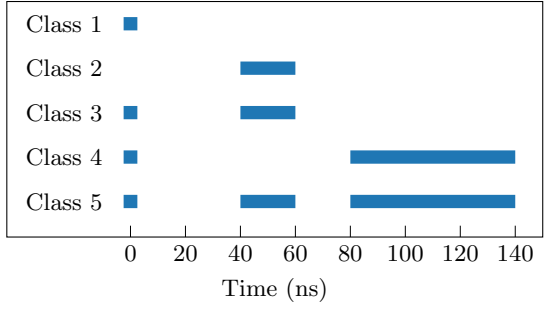
\begin{figure}[htbp]
\centering
\begin{tikzpicture}
\begin{axis}[
scale only axis,
width=0.82\columnwidth,
height=0.36\columnwidth,
xmin=-45, xmax=150,
ymin=0.2, ymax=5.5,
xlabel={Time (ns)},
xtick={0,20,40,60,80,100,120,140},
xtick pos=bottom,
ytick=\empty,
font=\fontsize{9}{11}\selectfont,
axis line style={black},
tick style={black},
grid=none
]

\node[anchor=east] at (axis cs:-7,5) {Class 1};
\node[anchor=east] at (axis cs:-7,4) {Class 2};
\node[anchor=east] at (axis cs:-7,3) {Class 3};
\node[anchor=east] at (axis cs:-7,2) {Class 4};
\node[anchor=east] at (axis cs:-7,1) {Class 5};

\definecolor{myblue}{RGB}{31,119,180}

% Class 1
\addplot[myblue, line width=5pt, line cap=butt, mark=none] coordinates {(-2.5,5) (2.5,5)};

% Class 2
\addplot[myblue, line width=5pt, line cap=butt, mark=none] coordinates {(40,4) (60,4)};

% Class 3
\addplot[myblue, line width=5pt, line cap=butt, mark=none] coordinates {(-2.5,3) (2.5,3)};
\addplot[myblue, line width=5pt, line cap=butt, mark=none] coordinates {(40,3) (60,3)};

% Class 4
\addplot[myblue, line width=5pt, line cap=butt, mark=none] coordinates {(-2.5,2) (2.5,2)};
\addplot[myblue, line width=5pt, line cap=butt, mark=none] coordinates {(80,2) (140,2)};

% Class 5
\addplot[myblue, line width=5pt, line cap=butt, mark=none] coordinates {(-2.5,1) (2.5,1)};
\addplot[myblue, line width=5pt, line cap=butt, mark=none] coordinates {(40,1) (60,1)};
\addplot[myblue, line width=5pt, line cap=butt, mark=none] coordinates {(80,1) (140,1)};

\end{axis}
\end{tikzpicture}
\caption{Event classification scheme based on pulse occurrence within predefined time windows.}
\label{fig:event_classification}
\end{figure}

To help understand the pulse waveforms that are detected in the FastFrame acquisition readout, we create a pulse event classification methodology to bin each detected pulse based on predefined time window characteristics. All measurements were performed using a FastFrame acquisition of 1000 cycles, with a trigger threshold set at 4.9 mV for all channels. A pulse was defined as a valid signal when the recorded voltage exceeded this threshold. Pulses were classified according to their occurrence within three predefined time windows, illustrated in Fig.~\ref{fig:event_classification}.

\begin{table}[H]
\centering
\begin{tabular}{|c|c|c|c|c|c|}
\hline
& Class 1 & Class 2 & Class 3 & Class 4 & Class 5\\
\hline
CH1 (Pad A1) & 262 & 21 & 17 & 34 & 4 \\ 
\hline
CH2 (Pad A2) & 331 & 23 & 12 & 39 & 2 \\ 
\hline
CH3 (Pad B1)  & 388 & 24 & 46 & 90 & 21 \\ 
\hline
CH4 (Pad B2) & 442 & 35 & 70 & 127 & 25 \\ 
\hline
\end{tabular}
\caption{Distribution of event classes for channels CH1 to CH4, which recorded pixel pads A1, A2, B1, and B2 based on predefined timing windows.}
\label{tab:eventdist}
\end{table}

\begin{table*}[ht!]
% \centering
\begin{tabular}{|c||c||c|c||c|c||c|c|c|c|c|c|c|c|c|c|c|c|}
\hline
 & P.C. 
 & \multicolumn{2}{c||}{Top MCP} 
 & \multicolumn{2}{c||}{Bottom MCP} 
 & \multicolumn{6}{c|}{Time Delay (ns)} 
 \\
\hline
Set & $V^{PC}$ & $V_T^T$ & $V_B^T$ & $V_T^B$ & $V_B^B$ & A (Mean) & A (SD) & A (SEM) & B (Mean) & B (SD) & B (SEM) \\
\hline
1 & 2200 & 2150 & 1275 & 1075 & 200 & 109.59 & 14.89 & 0.95 & 109.53 & 15.05 & 0.66 \\
2 & 2200 & 2100 & 1250 & 1050 & 200 & 110.53 & 17.10 & 1.03 & 109.76 & 15.14 & 0.64 \\
3 & 2200 & 2075 & 1275 & 1075 & 200 & 108.68 & 13.82 & 0.97 & 110.17 & 15.79 & 0.68 \\
4 & 2200 & 2050 & 1200 & 1000 & 200 & 108.48 & 15.61 & 1.38 & 111.22 & 14.09 & 0.78 \\
5 & 2200 & 2000 & 1200 & 1000 & 200 & 106.96 & 14.57 & 1.61 & 111.63 & 14.89 & 0.85 \\
\hline
\end{tabular}
\caption{Time delay analysis of the second pulse peak under different operating voltage configurations (N=1000). }
\label{tab:3}
\end{table*}

Using this classification scheme, data was collected from pads A1, A2, B1 and B2 as seen in Table~\ref{tab:eventdist}. During the 1000 recorded cycles, Class 1 events were observed to dominate the dataset, while events involving coincident pulses in multiple time windows, such as Class 3, 4, and 5, occurred much less frequently. This indicates that multi window coincidence events are intrinsically low probability processes under the tested conditions. A clear asymmetry was also observed between the channels: CH3 and CH4, corresponding to pad B1 and B2, consistently exhibited higher event counts in all classes compared to CH1 and CH2 (pad A1 and A2).\\
\indent In a total of 1000 measurements, when a tail structure was present, the probability of observing an additional peak around $\sim$60~ns or $\sim$110~ns was 2.55\% for A pads and 8.325\% for B pads.
For events in which both time periods that peaks were simultaneously present, the average probability was 0.3\% for A pads and 2.3\% for B pads.
These results indicate that delayed secondary structures occur more frequently in B pads than in A pads under otherwise comparable conditions. This difference may be partially influenced by cross talk originating from the A side.
%Based on Table III
Initially we suspect the wide time distribution of the second delayed peak may be associated with an ion feedback mechanism within the MCP structure. But we observe that the amplitude of the delayed peaks does not increase significantly with higher operating voltage in table \ref{tab:3}. If the delayed structure were entirely dominated by ion feedback, one would typically expect a stronger voltage dependence, since higher voltages increase electron energy and ionization probability. The relatively weak voltage dependence observed suggests that the ion feedback process may be limited or influenced by additional factors that require further investigation.

\section{2D LAPPD Simulation}

To investigate the cause of additional pulses observed in the FastFrame experimental data, we have built a 2D simulation model of the LAPPD Gen 2 coded in Python. This model takes a uniform electric field to account for the voltages at each MCP and the in-between locations. We start by assuming that 1 or more photoelectrons (PE) start at the photocathode, that are then driven toward the MCPs via the electric field, and then spread out from the initial MCP to many MCP channels at the second layer. When an electron makes contact with one of the MCP channel, additional electrons are spawned and recoil from the impact on the chevron. Each additional electron produced can continue to interact with the channels until they exit the MCP. This effectively reproduces the electron avalanche effect, and the process continues until all electrons reach the anode. We record the radial position relative to the center of one of the LAPPD pixels, as well as the time of arrival upon its termination at the anode. This produces a basic model that takes a photoelectron's position in time and space to simulate a voltage signal that one would measure in a real LAPPD experiment. 

\subsection{Algorithm}
We determine the starting position and energy for a photoelectron using simply two gaussian distributions. From those distributions and Equations \ref{eq:v_r},\ref{eq:v_z} we find can find the initial velocities in the $\hat{r}$ and $\hat{z}$ directions:
\begin{equation}
        v_{r,0}=\sqrt{\frac{2E_0e}{m_e}}\sin \theta
    \label{eq:v_r}
\end{equation}
\begin{equation}
    v_{z,0}=\sqrt{\frac{2E_0e}{m_e}}\cos \theta  
    \label{eq:v_z}
\end{equation}

$E_0$ is the initial energy of the PE in eVs, $e$ is the charge of an electron, $m_e$ is the mass of an electron, and $\theta$ is the angle between the $r$ and $z$ planes. In cylindrical coordinates, neglecting edge effects (a valid approximation away from the LAPPD perimeter), we can determine the time of flight $t$ for an electron traversing a distance $z$ (from MCP exit to anode) by solving: 
\begin{equation}
    z = v_0^\parallel t + \frac{1}{2} \frac{e E}{m_e}t^2
\end{equation}
The corresponding radial displacement on the anode is then: 
\begin{equation}
    r = v_0^\perp t
\end{equation}
For collisions, we simply take the normal direction of the boundary. To determine the number of secondary emissions after a collision, we use Equation \ref{eq:delta} and a random Poisson distribution with $\delta$ as the mean.
\begin{equation}
    \delta=\delta_{max}\frac{E}{E_{max}}\exp{\Big({1-\frac{E}{E_{max}}}\Big)} \label{eq:delta}
\end{equation}
Where $\delta_{max}$ is a scalar to control secondary emission based on the material and geometry of the MCP, E is the energy of the Photoelectron in eV, and $E_{max} $ is the energy that generates the most secondary emissions. We fit these parameters in such a way to have a gain on the order of $10^{6}-10^{7}$. When generating a voltage signal we simply take the radial position and fit each point to a gaussian to approximately determine the relative strength of the signal from arriving electron. We note that in Figure \ref{fig:back_scattering} the distribution of electrons are not gaussian, but the vast majority of electron are grouped near the center where a gaussian can represent fairly well. %Might need to reword...(James)
So an electron closer to the center ($r=0$) will have a stronger signal compared to one farther out. %To James: Reviewers might have issues with this and want more clarity "why does the signal strength decrease from the center of the radius?"
Then we bin these electrons based on the time each electron arrived and those determine the current which we can then use to find a voltage. 

To ensure computational efficiency, we model a slice of the LAPPD away from the walls so the path trajectory is simple to model. Based on the chevron channel pattern, we expect there to be a bias of electrons angled toward the left as they move towards the anode. All secondary electrons have an average energy of about 2 eV when generated. We did not consider coulomb interactions with the reasoning that these electrons are in a strong electric field and the trajectories would not be significantly affected. Incorporating coulomb interactions would also significantly increase computational overhead. 

\subsection{Initial Simulation Results}

\FloatBarrier
    \begin{figure}
        \begin{overpic}[width=1\linewidth]{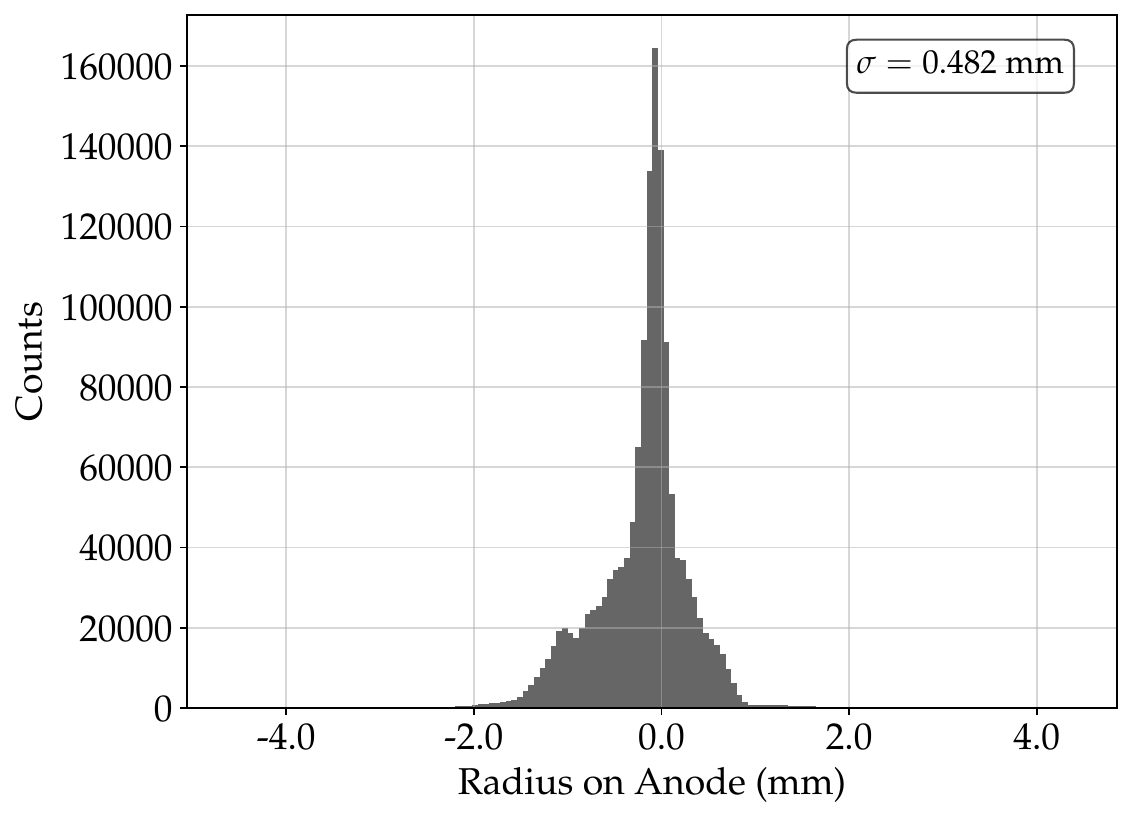}
        \end{overpic}
        \centering
        \caption{The electron cloud spread seen generally. Note that the spread of electrons is a bit to the left of center with a decay near the edges. This is due to the geometry of the LAPPD as the patterning has a slight bias to the left.}
        \label{fig:back_scattering}

        % Fix spacing with legend 
    \end{figure}

    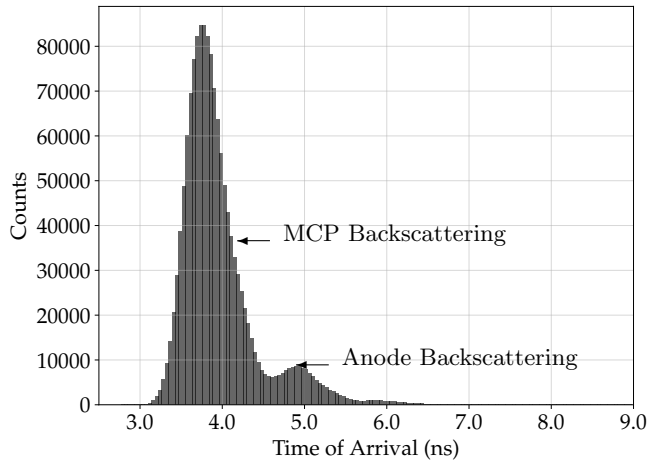
\begin{figure}
        \centering
        \begin{overpic}[width=1\linewidth]%, grid,tics=5]
        {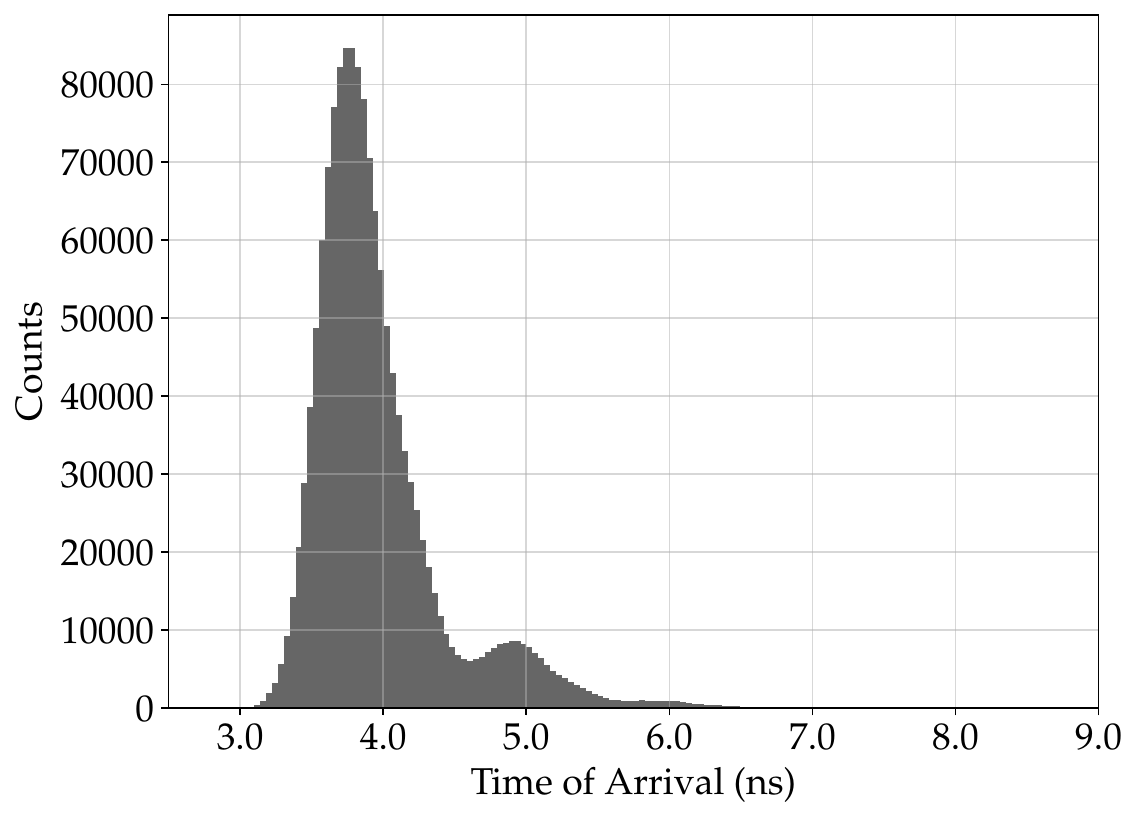}
    \put(41, 35){\color{black}\vector(-1,0){5}}
    \put(43, 35){MCP Backscattering}
    \put(50, 16){\color{black}\vector(-1,0){5}}
    \put(52, 16){Anode Backscattering}
        \end{overpic}
        \caption{Corresponding timing distribution for Figure \ref{fig:back_scattering} where we note there are three gaussian distributions shown. The first being the main SPE signal. The second signal is very well mixed with the SPE signal around 4.2~ns. Then the last signal at 5~ns which originates from anode back scattering.}
        \label{fig:time_scattering}
    \end{figure}
    
To ensure consistency with experimental data, we perform a sanity check by seeing if the model reproduces SPE pulses like those observed in the LAPPD. As mentioned in the algorithm, we first record the radial position ($n \sim 1 \times 10^6$ electrons) and observe a slight bias to the left in Fig.~\ref{fig:back_scattering}. This agrees with our expectations due to the chevron channel pattern. 
Within the time domain, we manage to reproduce the expected SPE pulse at a time of arrival of $\sim$4 ns to the anode. We notice additional signals after the initial expected reproduced SPE pulse with a small signal 1~ns after the main SPE pulse and potentially another gaussian like signal mixed into the SPE signal, seen in Fig.~\ref{fig:time_scattering}. 
We believe that these are a result of anode and MCP back-scattering, and will investigate this further in the model.

Initially it seems reasonable that the spread of the electron cloud does not lead to a major source of cross-talk that we see in experiment. The pixels are 25.4~mm by 25.4~mm in size, while the spread is on the order of a couple millimeters. It is more likely that cross-talk is a feature coming from the coupling of the pixels. However, if the incoming particle were to hit near the edge between two pixels, cross-talk effects from the electron cloud may become significant.

% \FloatBarrier

\begin{figure}
\begin{subfigure}
    \centering
    \includegraphics[width=1\linewidth]{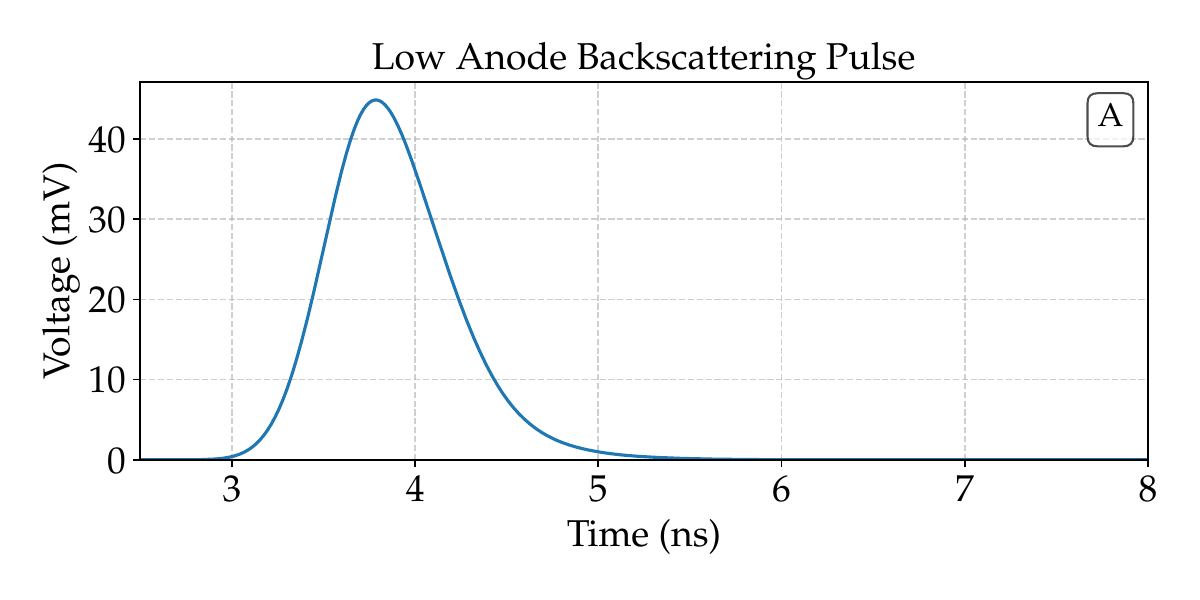}
\end{subfigure}
\begin{subfigure}
    \centering
    \includegraphics[width=1\linewidth]{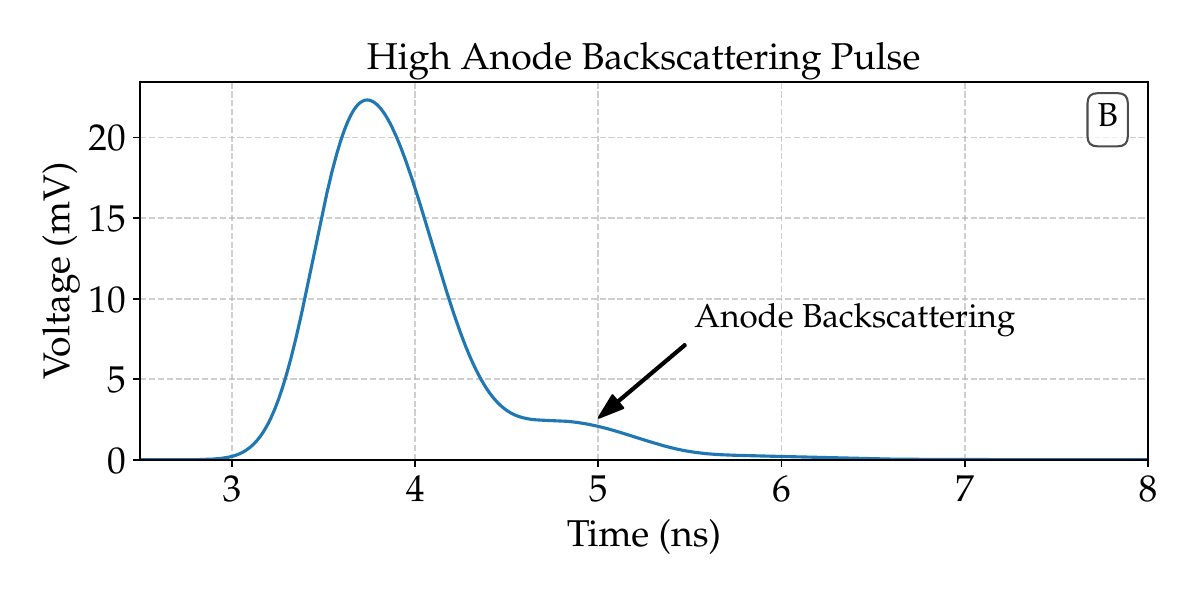}

\end{subfigure}
\caption{Comparison of differing levels of anode back-scattering. (A) We notice that the signal appears to have the main SPE pulse with an extended tail. (B) We can see the primary SPE pulse however there exists a notable signal need the end of the SPE pulse. This signal primary originates from anode back-scattering. However in both (A) and (B) it is not apparent that MCP back-scattering affects the signal.}
\label{fig:back_anode_scattering}
\end{figure}

\subsection{MCP and Anode Back-scattering}
We also consider anode and MCP back-scattering of electrons as an inherent feature that arises from the nature of the simulation. MCP back-scattering occurs in two phases. One is when the initial photoelectron hits in between the channels of the MCP and rebounds. This causes the signal to be delayed and come later than expected. In simulation, this is something we directly see as sometimes the signal arrives around 5~ns as opposed to 3~ns. The second phase is when secondary electrons miss the bottom MCP and rebound back into the top MCP, causing an additional mini electron avalanche. 

From our preliminary results, it seems that MCP back-scattering is very faint and often is mixed into the signal itself. If we were to add noise, there would be no reasonable way to distinguish the two. 
Anode back-scattering is simply electrons that rebound from the anode and are collected by the anode some time later. Without noise, we can sometimes see these additional signals. In real experiments with noise, these back-scattering manifest as a longer tail in the main signal as seen in Figure \ref{fig:back_anode_scattering}. 

% \FloatBarrier
\subsection{Ion After-pulses}
\begin{figure}
\begin{subfigure}
    \centering
    \includegraphics[width=1\linewidth]{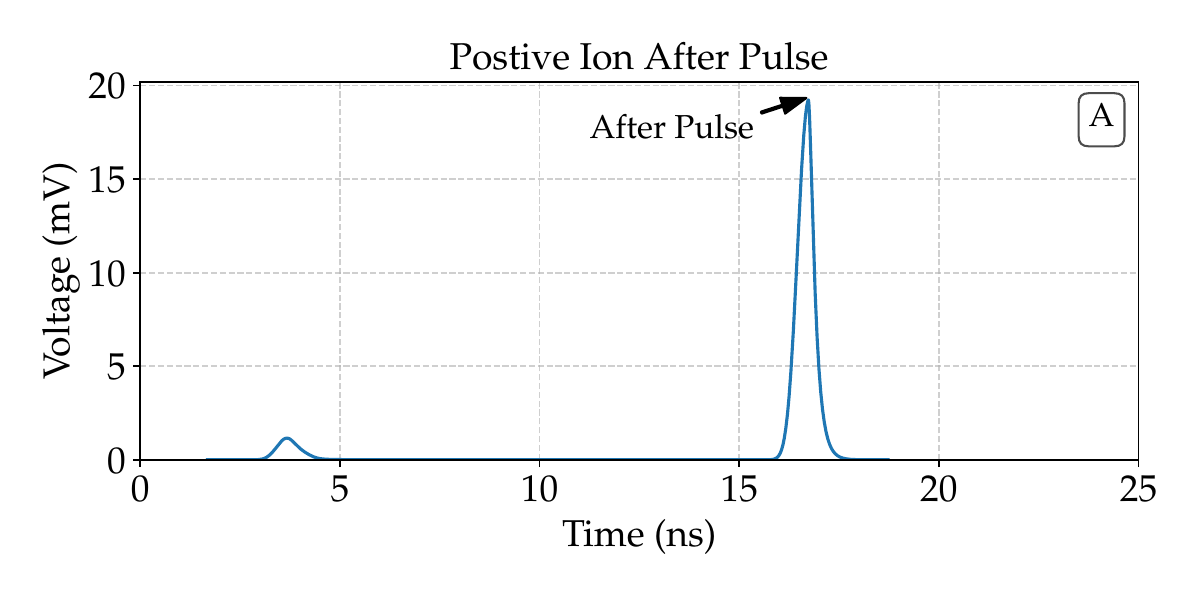}
\end{subfigure}
\begin{subfigure}
    \centering
    \includegraphics[width=1\linewidth]{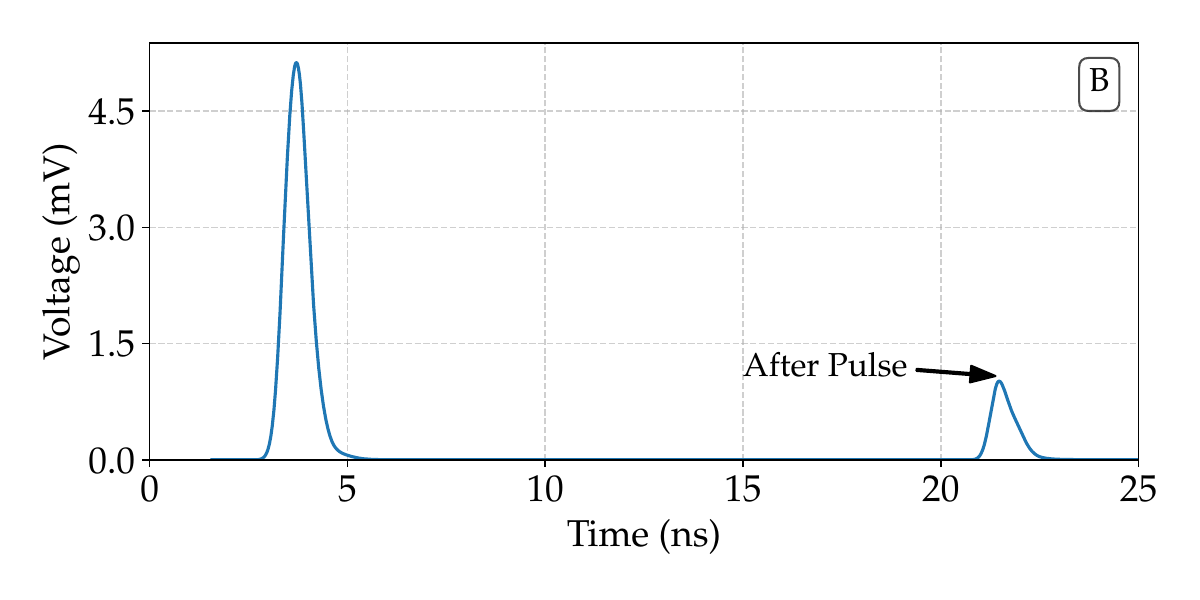}

\end{subfigure}
\caption{Comparison of two different cases when ion after-pulses are generated. (A) We can notice that the after-pulse signal is strong relative to the SPE pulse. (B) The opposite case of A, we notice that the after-pulse signal is small relative to the SPE pulse. Considering noise, we would most likely only detect the after-pulse in case A when it is caught by data acquisition trigger settings.}
\label{fig:ion_afterpulse}
\end{figure}

Within our model, we have also included ion-feedback by using a combination of the impact energy from electrons and a probability distribution to "spawn" protons. To model the number of photoelectrons emitted we simply use the following equation:~\ref{eq:Photoelectron_effect}. 
\begin{equation}
    E_k = E_{ion} -\Phi
    \label{eq:Photoelectron_effect}
\end{equation}
These ions are spawned near the entry of the top MCP and occasionally spawn part-way inside the MCP. As a result, we find that after-pulses appear around 15-30~ns from the start of the simulation. We notice from Figure~\ref{fig:ion_afterpulse} that the after-pulse can be weaker or stronger than the SPE pulse and occurs due to the release of multiple photoelectrons when the ion makes contact with the photocathode. Since the number of generated photoelectrons and ions is capped to one (to keep simulation runtimes fast), it is possible that the after-pulse signal is stronger than the SPE pulse in reality for certain scenarios. We note that this could potentially be a source of error and confusion if the after-pulse is triggered upon and mistaken as the initial SPE event. We can also consider that the system is quite sensitive to first multiplication stages. These simulated pulses seem to align with pulses defined in the event classification scheme, specifically Cases 1 to 3 shown in Figure \ref{fig:event_classification}. 

In the FastFrame raw data capture seen in Fig.~\ref{fig:FastFrame}, there are two regions that could have resulted from ion feedback. We note that if they are positive ions, they would likely have to appear before the top MCP. Otherwise, there would be secondary emissions while the positive ion travels through the detector and would likely generate fairly large signals. More research is needed on this behavior as it does not seem that positive ion feedback alone can explain the full behavior of the additional peaks at 60~ns and 110~ns.

\section{Conclusion}

In this follow-on study to our previous publication~\cite{Slava2025_RSI}, we investigate key performance characteristics of LAPPD Gen 2 and develop a comprehensive device response model. We report several previously under-characterized behaviors with direct implications for experimental operation and calibration. In particular, we quantify intrinsic channel-to-channel crosstalk and show how it can bias timing and position reconstruction if not explicitly modeled. We also observed elevated dark count rates that could be due to increase thermal emissions when changing voltage settings. We further examine the potential roles of electron backscatter and ion feedback, which introduce secondary signals that can degrade event reconstruction. This is especially true in low-light, sparse-photon environments relevant to next-generation neutrino detectors such as Theia and KamLAND2. Looking ahead, FPGA-resident machine-learning inference may provide an efficient route to real-time feature extraction and data reduction for LAPPD readout, enabling “smart” photosensor operation.

In practical operation, we note that optimizing for a single performance metric, such as dark-count rates, comes at the expense of other LAPPD characteristics, primarily gain and pixel timing resolution. As a result, no single operating point is universally optimal; the relative importance of each parameter depends on the intended application, and so optimization involves user-defined trade-offs. For example, optimizing voltage settings to minimize the dark-count rate may be less critical for applications focused on scintillation detection, while pixel timing resolution may be prioritized in dual Cherenkov-scintillation detection.

We note that a recently published study has developed a similar simulation model for the LAPPD Gen 2 \cite{Korpar:2025rcj} to explain photoelectron propagation and resistive anode readout within the LAPPD. Their work has developed analytical models by focusing on the charge deposited in a region of pixels to compare with experiment. Compared to our study, we worked in a 2D case where we used Monte Carlo simulations to look at the secondary electron distribution onto the resistive anode. We also go into how the processed signal is affected by conditions within the detector without considering the coupling of the pixels, and attempt to explain sources of after-pulses in the LAPPD. We hypothesize that those after pulses result from positive ions, although there are additional unknown mechanisms that influence these after-pulses that can be the focus of further investigation.
%\newpage
% \section{Potential journals}

% AIP Advances

% APS Open Science

% Nuclear Instrumentation and Methods

% Journal of Instrumentations

\section{Acknowledgments}
We thank Mark A. Popecki from Incom for continuing technical support and useful discussion during this study.
We thank Bryan~Bowman and Karl~Gundal from Tektronix for their continuing support. We acknowledge Tomi Akindele for leading the larger project within which this research was conducted.
This research was supported in part by an appointment to the National Nuclear Security Administration Minority Serving Institutions Internship Program (NNSA-MSIIP), sponsored by the U.S. Department of Energy and administered by the Oak Ridge Institute for Science and Education.
The work of J.F. and S.W.S. was partially supported by DOE’s Reaching New Energy Workforce Initiative.
This work is supported by the U.S. Department of Energy National Nuclear Security Administration and Lawrence Livermore National Laboratory [Contract No. DE-AC52-07NA27344, release number LLNL-JRNL-2020218].

\bibliography{ref}

@article{Bhattacharya:2023nmj,
    author = "Bhattacharya, Deb Sankar and others",
    title = "{Characterization of LAPPD timing at CERN PS testbeam}",
    eprint = "2309.15011",
    archivePrefix = "arXiv",
    primaryClass = "physics.ins-det",
    doi = "10.1016/j.nima.2023.168937",
    journal = "Nucl. Instrum. Meth. A",
    volume = "1058",
    pages = "168937",
    year = "2024"
}

@article{Korpar:2025rcj,
    author = "Korpar, S. and Dolenec, R. and Grijalva, F. and Lozar, A. and Kodri{\v{c}}, A. and Kri{\v{z}}an, P. and Parashari, S. and Pestotnik, R. and Seljak, A. and {\v{Z}}ontar, D.",
    title = "{Characterisation of the LAPPD, a large area microchannel-plate PMT}",
    eprint = "2512.02990",
    archivePrefix = "arXiv",
    primaryClass = "physics.ins-det",
    month = "12",
    year = "2025"
}

@article{Slava2025_RSI,
    author = {Li, V. A. and Akindele, O. A. and Bondin, M. and Durham, S. R. and Foot, J. A. and Ford, M. J. and Stradleigh, S.-W.},
    title = {Initial assessment of second generation of large-area picosecond photodetectors with multi-channel systems-on-a-chip readout},
    journal = {Review of Scientific Instruments},
    volume = {96},
    number = {11},
    pages = {113102},
    year = {2025},
    month = {11},
    issn = {0034-6748},
    doi = {10.1063/5.0269181},
    url = {https://doi.org/10.1063/5.0269181}
}

@article{Yeh:2011zz,
    author = {M. Yeh and S. Hans and W. Beriguete and R. Rosero and L. Hu and R. L. Hahn and M. V. Diwan and D. E. Jaffe and S. H. Kettell and L. Littenberg},
    title = "{A new water-based liquid scintillator and potential applications}",
    journal = "Nucl. Instrum. Meth. A",
    volume = "660",
    pages = "51--56",
    year = "2011"
}

@article{Aberle_2014,
year = {2014},
month = {jun},
publisher = {},
volume = {9},
number = {06},
pages = {P06012},
author = {Aberle, C and Elagin, A and Frisch, H J and Wetstein, M and Winslow, L},
title = {Measuring directionality in double-beta decay and neutrino interactions with kiloton-scale scintillation detectors},
journal = {JINST}
}

@article{Caravaca_2020,
    author = {Caravaca, J. and Land, B.J. and Yeh, M. and Orebi Gann, G.D},
    title = {Characterization of water-based liquid scintillator for Cherenkov and scintillation separation},
    journal = {The European Physical Journal C},
    volume = {80},
    number = {867},
    year = {2020}
}

@article{anniecollab2024,
    author = {M. Ascencio-Sosa and Z. Bagdasarian and J. F. Beacom and M. Bergevin and M. Breisch and G. Caceres Vera and S. Dazeley and S. Doran and E. Drakopoulou and S. Edayath and others},
    title = "{Deployment of Water-based Liquid Scintillator in the Accelerator Neutrino Neutron Interaction Experiment}",
    journal = "JINST",
    volume = "19",
    number = "05",
    pages = "P05070",
    year = "2024"
}

@article{Kaptanoglu_2022,
    author = {Kaptanoglu, T. and Callaghan, E.J. and Yeh, M. and Orebi Gann, G.D.},
    title = "{Cherenkov and scintillation separation in water-based liquid scintillator using an LAPPD}",
    journal = "The European Physical Journal C",
    volume = "82",
    number = "169",
    year = "2022"
}

@article{Askins,
    author = {M. Askins and Z. Bagdasarian and N. Barros and E.W. Beier and E. Blucher and R. Bonventre and E. Callaghan and J. Caravaca and M. Diwan and S.T. Dye and others},
    title = {Theia: an advanced oiptical neutrino detector},
    journal = {The European Physical Journal C},
    volume = {80},
    number = {416},
    year = {2020}
}

@article{Goodrich1962,
    author = {Goodrich, G. W. and Wiley, W. C.},
    title = {Continuous Channel Electron Multiplier},
    journal = {Review of Scientific Instruments},
    volume = {33},
    number = {7},
    pages = {761-762},
    year = {1962},
    month = {07},
    issn = {0034-6748},
    doi = {10.1063/1.1717958},
    url = {https://doi.org/10.1063/1.1717958}
}

@book{Bell1973,
    author = {Bell, R. L.},
    title = {Negative electron affinity devices},
    publisher = {Claredon Press, Oxford},
    year = {1973}
}

@manual{Hamamatsu2017,
    author = {Hamamatsu},
    title = {Photomultiplier Tubes, Basics and Applications, Fourth Edition},
    year = {2017},
}

@ARTICLE{Morton1967,
  author={Morton, G. A. and Smith, H. M. and Wasserman, R.},
  journal={IEEE Transactions on Nuclear Science}, 
  title={Afterpulses in Photomultipliers}, 
  year={1967},
  volume={14},
  number={1},
  pages={443-448},
  doi={10.1109/TNS.1967.4324452}}

@article{STAUBERT1970,
title = {Possible effects of photomultiplier-afterpulses on scintillation counter measurements},
journal = {Nuclear Instruments and Methods},
volume = {84},
number = {2},
pages = {297-300},
year = {1970},
issn = {0029-554X},
doi = {https://doi.org/10.1016/0029-554X(70)90276-4},
url = {https://www.sciencedirect.com/science/article/pii/0029554X70902764},
author = {R. Staubert and E. Böhm and K. Hein and K. Sauerland and J. Trümper}
}

@article{HALL1973,
title = {Reduction of afterpulsing in a photomultiplier},
journal = {Nuclear Instruments and Methods},
volume = {112},
number = {3},
pages = {545-549},
year = {1973},
issn = {0029-554X},
doi = {https://doi.org/10.1016/0029-554X(73)90176-6},
url = {https://www.sciencedirect.com/science/article/pii/0029554X73901766},
author = {S.J. Hall and J. McKeown}
}

@article{AKCHURIN2007,
title = {A study on ion initiated photomultiplier afterpulses},
journal = {Nuclear Instruments and Methods in Physics Research Section A: Accelerators, Spectrometers, Detectors and Associated Equipment},
volume = {574},
number = {1},
pages = {121-126},
year = {2007},
issn = {0168-9002},
doi = {https://doi.org/10.1016/j.nima.2007.01.093},
url = {https://www.sciencedirect.com/science/article/pii/S0168900207001532},
author = {Nural Akchurin and Heejong Kim},
}

\section{Appendix}

\begin{table}[h]
\centering

\begin{tabularx}{\columnwidth}{|c||*{8}{>{\centering\arraybackslash}X|}}
\hline
[mm] & 1 & 2 & 3 & 4 & 5 & 6 & 7 & 8 \\
\hline
\hline
A & 24 & 40 & 67 & 91 & 91 & 66 & 37 & 23 \\
\hline
B & 4 & 49 & 86 & 102 & 104 & 89 & 48 & 28 \\
\hline
C & 22 & 60 & 80 & 108 & 82 & 80 & 60 & 25 \\
\hline
D & 25 & 59 & 68 & 111 & 111 & 77 & 62 & 26 \\
\hline
E & 35 & 50 & 85 & 102 & 104 & 86 & 55 & 35 \\
\hline
F & 32 & 50 & 89 & 89 & 84 & 92 & 50 & 35 \\
\hline
G & 37 & 45 & 61 & 103 & 93 & 65 & 54 & 28 \\
\hline
H & 16 & 51 & 69 & 92 & 94 & 74 & 56 & 11 \\
\hline
\end{tabularx}

\caption{Trace-length distances in millimeters to each unique pixel location. We note that these differences cause a measurable timing difference when computing two-pixel waveform delay statistics as seen in previous work \cite{Slava2025_RSI}, and must be accounted for.}

\label{tab:example}
\end{table}

\end{document}